\documentclass[pra,preprint,showpacs,amsmath,amssymb,nofootinbib]{revtex4}

\usepackage{amsthm}

\begin{document}

\newtheorem{theorem}{Theorem}
\newtheorem{lemma}{Lemma}

\newcommand\ket[1]{\ensuremath{|#1\rangle}}
\newcommand\bra[1]{\ensuremath{\langle#1|}}
\newcommand\iprod[2]{\ensuremath{\langle#1|#2\rangle}}
\newcommand\oprod[2]{\ensuremath{|#1\rangle\langle#2|}}

\title{Optimal conclusive discrimination of two states can be achieved locally}

\author{Zhengfeng Ji}
  \email{jizhengfeng98@mails.tsinghua.edu.cn}
\author{Hongen Cao}
  \email{cge01@mails.tsinghua.edu.cn}
\author{Mingsheng Ying}
\affiliation{
State Key Laboratory of Intelligent Technology and Systems, Department of Computer Science and Technology,\\
Tsinghua University, Beijing 100084, China
}

\date{\today}

\begin{abstract}
This paper constructs a LOCC protocol that achieves the global optimality in conclusive discrimination of any two states with arbitrary \textit{a priori} probability. This can be interpreted that there is no ``non-locality'' in the conclusive discrimination of two multipartite states.
\end{abstract}


\maketitle

One of the most fundamental different aspects between quantum and classical information processing is that, unlike the classical case, states of a quantum system are not always distinguishable. Only orthogonal states can be distinguished with certainty even if global operations are allowed. And when operations are restricted to local operation and classical communication (LOCC) only, there exists a set of globally distinguishable \textit{product} states that can not be identified locally \cite{BDF+99}. Surprisingly however, Walgate \textit{et al.}\ \cite{WSHV00} proved that any \textit{two} orthogonal multipartite states, entangled or not, can be distinguished perfectly using only LOCC. This result leads us to conjecture that LOCC is strong enough to perform optimal discrimination between any two states, generally non-orthogonal.

Two types of discrimination are usually considered for non-orthogonal states in the literature: one is conclusive discrimination and the other is the inconclusive one. The main difference between them is that the former allows the ``don't know'' claim but no wrong answers while the latter permits incorrect judgement of the system focusing on minimizing the probability of making errors. Virmani \textit{et al.}\ proved the above conjecture in the sense of inconclusive discrimination in Ref.~\cite{VSPM01} where they also confirmed the conjecture for a special class of states that are \textit{Schmidt correlated}. Chen \textit{et al.}\ \cite{CY01} considered the problem of distinguishing any two product states with arbitrary \textit{a priori} probability and confirmed the conjecture. Later, they proved it for all states with equal prior probability \cite{CY02}. In this paper, we give a distinguishing protocol that uses only LOCC and achieves the global optimality for any two states with arbitrary \textit{a priori} probability. This protocol completely solved the problem and gave a positive answer to the conjecture. That is, there is no ``non-locality'' in the conclusive discrimination of any two multipartite states.

As we are actually trying to prove that LOCC can perform discrimination of two states as efficiently as global operation, it is necessary to review the result of the global case first. The problem of identifying two states using global measurements has been considered by Ivanovic \cite{Ivan87}, Dieks \cite{Diek88}, Peres \cite{Pere88}, Jaeger and Shimony \cite{JS95} and we restate the general result as follows. Suppose the state is prepared as $\ket{\phi}$ with probability $s$ and $\ket{\psi}$ with probability $t$ where $s+t=1$ and $s\le t$, then the maximal probability of success is
\begin{equation}\label{eq:linear}
P_{max} = 1-2\sqrt{st}|\iprod{\phi}{\psi}|
\end{equation}
if $\sqrt{s/r} \ge |\iprod{\phi}{\psi}|$ and
\begin{equation}\label{eq:quadratic}
P_{max} = t(1-|\iprod{\phi}{\psi}|^2)
\end{equation}
otherwise. Our task is then to achieve this maximal probability using LOCC only. First, we prove the case of
\begin{equation}
\sqrt{\frac{s}{r}} \ge |\iprod{\phi}{\psi}| \label{eq:lc}
\end{equation}
and then utilize this result to show that the other case also holds.

Many techniques and ideas used in this paper are directly borrowed from the proof in Ref.~\cite{CY02}. The following lemma is introduced there and we restate it without proof. As in Ref.~\cite{CY02}, this lemma is also useful here to construct the protocol step by step.

\begin{lemma}\label{lem:u}
Let $M$ be $2 \times 2$ matrix whose diagonal elements are real and
\begin{equation}
U = \begin{bmatrix}
\cos \theta & \sin \theta e^{-i\omega}\\
\sin \theta e^{i\omega} & -\cos \theta
\end{bmatrix}.
\end{equation}
There exists $\omega$ such that the diagonal elements of $UMU^{\dagger}$ are real and this property is independent of $\theta$.
\end{lemma}

It was proved in Ref.~\cite{WSHV00} that the two states, $\ket{\phi}$ and $\ket{\psi}$, can be expressed in the following form:
\begin{eqnarray}
\ket{\phi} & = & \sum_{i=1}^{n}\sqrt{s_i}\ket{e_i}_A\otimes\ket{\eta_i}_B\nonumber\\
\ket{\psi} & = & \sum_{i=1}^{n}\sqrt{t_i}\ket{e_i}_A\otimes\ket{\gamma_i}_B\label{eq:eipe}
\end{eqnarray}
satisfying 
\begin{equation}
\sqrt{s_i t_i}\iprod{\eta_i}{\gamma_i}_B = \sqrt{s_j t_j}\iprod{\eta_j}{\gamma_j}_B
\end{equation}
where $\{\ket{e_i}\}$ forms an orthonormal basis set. Walgate \textit{et al.}\ utilized this result to show that $\iprod{\eta_i}{\gamma_i} = 0$ for all $i=1, 2, \cdots, n$ and proved that Alice and Bob can always distinguish between the two orthogonal states perfectly via LOCC operations. This expansion is also where we start.

As it's free to add a global phase to $\ket{\psi}$ in Eq.~(\ref{eq:eipe}), we can assume that $\iprod{\phi}{\psi} > 0$ in this paper without loss of generality. Our proof begins with the following lemma.

\begin{lemma}\label{lem:lee}
In a properly chosen orthonormal basis $\ket{i}$ on Alice's side, $\ket{\phi}$ and $\ket{\psi}$ can be expressed as:
\begin{eqnarray}
\ket{\phi} & = & \sum_{i=1}^n \sqrt{s_i}\ket{i}\ket{\eta_i}, \nonumber\\
\ket{\psi} & = & \sum_{i=1}^n \sqrt{t_i}\ket{i}\ket{\gamma_i}\label{eq:lee},
\end{eqnarray}
where $\iprod{\eta_i}{\gamma_i}$ is real\footnote{Depends on the assumption of $\iprod{\phi}{\psi} > 0$ we have made.} and $s s_i\le t t_i$ for all $i$.
\end{lemma}

Simple observation shows that this lemma is equivalent to the first theorem in Ref.~\cite{CY02}. Let $s=t$, we have $s_i\le t_i$ for all $i$ and thus $s_i = t_i$ which is just the first theorem in Ref.~\cite{CY02} since $\sum s_i=\sum t_i=1$. On the other hand, the theorem in Ref.~\cite{CY02} indeed gives an eligible expansion of Lemma~\ref{lem:lee} since $s t_i$ is always less than or equal to $t t_i$. In the following lemma, we further strengthen the expansion to make it work in the general case.

\begin{lemma}\label{lem:iple}
In a properly chosen orthonormal basis $\ket{i}$ on Alice's side, $\ket{\phi}$ and $\ket{\psi}$ can be expressed as:
\begin{eqnarray}
\ket{\phi} & = & \sum_{i=1}^n \sqrt{s_i}\ket{i}\ket{\eta_i}, \nonumber\\
\ket{\psi} & = & \sum_{i=1}^n \sqrt{t_i}\ket{i}\ket{\gamma_i}\label{eq:iple},
\end{eqnarray}
where $\iprod{\eta_i}{\gamma_i}$ is real, $s s_i\le t t_i$ and
\begin{equation*}
\sqrt{\frac{ss_i}{tt_i}} \ge \iprod{\eta_i}{\gamma_i}
\end{equation*}
for $i=1, 2, \cdots, n$.
\end{lemma}

\begin{proof}
We begin with Eq.~(\ref{eq:lee}). If there is some $i$, say $1$, such that
\begin{equation}
\sqrt{\frac{ss_1}{tt_1}} < \iprod{\eta_1}{\gamma_1},\label{eq:nl}
\end{equation}
then there must also exists some $j$, say $2$, satisfies that
\begin{equation*}
\sqrt{\frac{ss_2}{tt_2}} > \iprod{\eta_2}{\gamma_2}.
\end{equation*}
Otherwise we have
\begin{equation}
\sum_{i=1}^n \sqrt{\frac{s}{t}} s_i < \sum_{i=1}^n \sqrt{s_it_i} \iprod{\eta_i}{\gamma_i} = \iprod{\phi}{\psi}
\end{equation}
which contradicts with Eq.~(\ref{eq:lc}). Employing the scheme of changing the basis on Alice's side using unitary transformation $U$ defined in Lemma~\ref{lem:u} just as what has been done in Ref.~\cite{CY02}, we have the following relation \cite{HJW93}:
\begin{eqnarray}
\begin{pmatrix}
\sqrt{s_1^{'}}\ket{\eta_1^{'}} \\
\sqrt{s_2^{'}}\ket{\eta_2^{'}}
\end{pmatrix}
& = &
\begin{bmatrix}
\cos \theta & \sin \theta e^{-i\omega}\\
\sin \theta e^{i\omega} & -\cos \theta
\end{bmatrix}
\begin{pmatrix}
\sqrt{s_1}\ket{\eta_1} \\
\sqrt{s_2}\ket{\eta_2}
\end{pmatrix}\nonumber\\
\begin{pmatrix}
\sqrt{t_1^{'}}\ket{\gamma_1^{'}} \\
\sqrt{t_2^{'}}\ket{\gamma_2^{'}}
\end{pmatrix}
& = &
\begin{bmatrix}
\cos \theta & \sin \theta e^{-i\omega}\\
\sin \theta e^{i\omega} & -\cos \theta
\end{bmatrix}
\begin{pmatrix}
\sqrt{t_1}\ket{\gamma_1} \\
\sqrt{t_2}\ket{\gamma_2}
\end{pmatrix}
\end{eqnarray}
from which the modification in the expansion can be calculated as
\begin{subequations}\label{eq:np}
\begin{eqnarray}
s_1^{'} & = & s_1 \cos^2 \theta + s_2 \sin^2 \theta + x \sin \theta \cos \theta, \label{eq:npa}\\
s_2^{'} & = & s_2 \cos^2 \theta + s_1 \sin^2 \theta - x \sin \theta \cos \theta, \label{eq:npb}\\
t_1^{'} & = & t_1 \cos^2 \theta + t_2 \sin^2 \theta + y \sin \theta \cos \theta, \label{eq:npc}\\
t_2^{'} & = & t_2 \cos^2 \theta + t_1 \sin^2 \theta - y \sin \theta \cos \theta, \label{eq:npd}
\end{eqnarray}
\end{subequations}
\begin{widetext}
\begin{subequations}\label{eq:nip}
\begin{eqnarray}
\sqrt{s_1^{'}t_1^{'}}\iprod{\eta_1^{'}}{\gamma_1^{'}} & = & \cos^2 \theta (\sqrt{s_1 t_1} \iprod{\eta_1}{\gamma_1}) + \sin^2 \theta (\sqrt{s_2 t_2} \iprod{\eta_2}{\gamma_2}) + z \cos \theta \sin \theta,\\
\sqrt{s_2^{'}t_2^{'}}\iprod{\eta_2^{'}}{\gamma_2^{'}} & = & \cos^2 \theta (\sqrt{s_2 t_2} \iprod{\eta_2}{\gamma_2}) + \sin^2 \theta (\sqrt{s_1 t_1} \iprod{\eta_1}{\gamma_1}) - z \cos \theta \sin \theta,
\end{eqnarray}
\end{subequations}
\end{widetext}
where
\begin{eqnarray*}
x & = & e^{-i\omega} \sqrt{s_1 s_2} \iprod{\eta_1}{\eta_2} + e^{i\omega} \sqrt{s_1 s_2} \iprod{\eta_2}{\eta_1},\\
y & = & e^{-i\omega} \sqrt{t_1 t_2} \iprod{\gamma_1}{\gamma_2} + e^{i\omega} \sqrt{t_1 t_2} \iprod{\gamma_2}{\gamma_1},\\
z & = & e^{-i\omega} \sqrt{s_1 t_2} \iprod{\eta_1}{\gamma_2} + e^{i\omega} \sqrt{s_2 t_1} \iprod{\eta_2}{\gamma_1}.
\end{eqnarray*}
We prove that it is always possible to get
\begin{equation}
\sqrt{\frac{ss_1^{'}}{tt_1^{'}}} = \iprod{\eta_1^{'}}{\gamma_1^{'}},\label{eq:lt}
\end{equation}
using the above change in basis. From Eq.~(\ref{eq:np}, \ref{eq:nip}), this equality is expanded as
\begin{widetext}
\begin{equation}
\sqrt{\frac{s}{t}} (s_1 \cos^2 \theta + s_2 \sin^2 \theta + x \sin \theta \cos \theta) = \cos^2 \theta (\sqrt{s_1 t_1} \iprod{\eta_1}{\gamma_1}) + \sin^2 \theta (\sqrt{s_2 t_2} \iprod{\eta_2}{\gamma_2}) + z \cos \theta \sin \theta,
\end{equation}
or equivalently
\begin{equation}
(\sqrt{\frac{s}{t}} s_2 - \sqrt{s_2 t_2} \iprod{\eta_2}{\gamma_2}) \tan^2 \theta + (\sqrt{\frac{s}{t}}x-z) \tan \theta + (\sqrt{\frac{s}{t}} s_1 - \sqrt{s_1 t_1} \iprod{\eta_1}{\gamma_1}) = 0.
\end{equation}
\end{widetext}
Since $\sqrt{\frac{s}{t}} s_2 - \sqrt{s_2 t_2} \iprod{\eta_2}{\gamma_2}$ and $\sqrt{\frac{s}{t}} s_1 - \sqrt{s_1 t_1} \iprod{\eta_1}{\gamma_1}$ differ in sign, the above quadratic equation of $\tan \theta$ has two real roots with different sign. After this change in basis, we still have $ss_1^{'} \le tt_1^{'}$ since
\begin{equation*}
\sqrt{\frac{ss_1^{'}}{tt_1^{'}}} = \iprod{\eta_1^{'}}{\gamma_1^{'}} \le 1.
\end{equation*}
Properly choose one of the two roots such that the sign of $\tan \theta$ is the same as the one of $sx-ty$, we can also have
\begin{eqnarray*}
ss_2^{'} & \le & ss_2 \cos^2 \theta + ss_1 \sin^2 \theta - sx \sin \theta \cos \theta\\
         & \le & tt_2 \cos^2 \theta + tt_1 \sin^2 \theta - ty \sin \theta \cos \theta\\
         & =   & tt_2^{'}
\end{eqnarray*}
Thus, if there are $i$ such that Eq.~(\ref{eq:nl}) holds, we can modify the basis properly to obtain Eq.~(\ref{eq:lt}) and keep the real property of the two inner-products involved. Although the sign of $\sqrt{\frac{ss_2^{'}}{tt_2^{'}}} - \iprod{\eta_2^{'}}{\gamma_2^{'}}$ is not known after such a change, we claim that the above procedure terminates in finite steps leaving a expansion that satisfies our requirement. In fact, we can evaluate the expansion of $n$ terms with the value of $2 \mathcal{E} + \mathcal{L}$ where $\mathcal{E}$ and $\mathcal{L}$ are number of terms that satisfy $\sqrt{\frac{ss_i}{tt_i}} = \iprod{\eta_i}{\gamma_i}$ and $\sqrt{\frac{ss_i}{tt_i}} > \iprod{\eta_i}{\gamma_i}$ respectively. Each step increases the evaluation by at least one but the evaluation has a maximal value of $2n$. Thus the procedure stops in finite steps, and then there must be no items of $\sqrt{\frac{ss_i}{tt_i}} < \iprod{\eta_i}{\gamma_i}$ otherwise the procedure can continue.
\end{proof}

Because of the strategy we will take is the same as the one in Ref.~\cite{CY02}, the expansion in Eq.~(\ref{eq:lee}) was further modified to a stronger version of nonnegative inner-products by introducing an auxiliary system $S$ just as what has been done in Ref.~\cite{CY02}. We restate it in the following lemma.

\begin{lemma}\label{lem:pipe}
There exists unitary transformation $U^{AS}$ on Alice's side and the auxiliary system such that $\ket{\phi}$ and $\ket{\psi}$ can be expressed as
\begin{eqnarray}
U^{AS}\ket{0}_S\ket{\phi} & = & \sum_{i=1}^m \sqrt{s_i} \ket{i}_S \ket{\phi_i} + \sum_{i=m+1}^N \sqrt{s_i} \ket{i}_S \ket{i} \ket{\eta_i}, \nonumber\\
U^{AS}\ket{0}_S\ket{\psi} & = & \sum_{i=1}^m \sqrt{t_i} \ket{i}_S \ket{\psi_i} + \sum_{i=m+1}^N \sqrt{t_i} \ket{i}_S \ket{i} \ket{\gamma_i},\label{eq:pipe}
\end{eqnarray}
where $\iprod{\phi_i}{\psi_i} = 0$ for $i\le m$ and
\begin{equation*}
1 \ge \sqrt{\frac{ss_i}{tt_i}} \ge \iprod{\eta_i}{\gamma_i} \ge 0
\end{equation*}
for $i > m$.
\end{lemma}

In Ref.~\cite{CY02}, unitary transformation $U_1^{AS}$ is chosen such that
\begin{eqnarray}
U_1^{AS}\ket{0}_S\ket{\phi} & = & \sqrt{s_1} \ket{\chi}_{AS} \ket{\eta_1} + \sqrt{s_2} \ket{\chi^{\perp}}_{AS} \ket{\eta_2} + \cdots\nonumber\\
U_1^{AS}\ket{0}_S\ket{\phi} & = & \sqrt{t_1} \ket{\chi}_{AS} \ket{\gamma_1} + \sqrt{t_2} \ket{\chi^{\perp}}_{AS} \ket{\gamma_2} + \cdots
\end{eqnarray}
Using the same unitary transformation and almost the same arguments, we are able to prove the above lemma. The only thing needs to be clarified here is that in Ref.~\cite{CY02}, the transformation is required to leave same probabilities to all corresponding term while in here, it is easy to see that the ratio of two corresponding probabilities does not change after the transformation which guarantees the requirements of our lemma.

Returning to the problem of distinguishing two states of the case $\sqrt{\frac{s}{t}} \ge \iprod{\phi}{\psi}$, we are now able to give the protocol by translating the expansion in Eq.~(\ref{eq:pipe}). Alice first performs the transformation $U^{AS}$ on both her side and the auxiliary system. Next, she measures the joint system on her side according to the states $\ket{i}_S$. If the outcome $i$ of the measurement is less than or equal to $m$, Alice and Bob can carry out the optimal discrimination with certainty as proved in \cite{WSHV00}. Otherwise, she tells the outcome to Bob who can then performs the optimal discrimination on his side.

The proof of global optimality comes as follows. Beware that before Bob's discrimination, he should first calculate the proper prior distribution probability since Alice's measurement may have changed them. Let $P_i$ be the probability of getting the measurement outcome $i$ on Alice's side and $P_{\ket{\phi}|i}$, $P_{\ket{\psi}|i}$ be conditional probability when outcome $i$ is observed. It's easy to see that $P_i = ss_i + tt_i$ and $P_{\ket{\phi}|i} = ss_i/P_i$, $P_{\ket{\psi}|i} = tt_i/P_i$. Thus the probability of successful discrimination is
\begin{eqnarray}
P_{max}^{LOCC} & = & \sum_{i=1}^m P_i + \sum_{i=m+1}^N P_i (1 - 2 \sqrt{ P_{\ket{\phi}|i} P_{\ket{\psi}|i} } \iprod{\eta_i}{\gamma_i})\nonumber\\
        & = & 1 - 2 \sum_{i=m+1}^N \sqrt{ ss_i tt_i } \iprod{\eta_i}{\gamma_i}\nonumber\\
        & = & 1 - 2 \sqrt{st} \iprod{\phi}{\psi}\nonumber\\
        & = & P_{max}
\end{eqnarray}

Hoping to prove the case of $\sqrt{\frac{s}{t}} < \iprod{\phi}{\psi}$, Alice and Bob need only to apply the protocol assuming that their system is prepared with probability $s^{*}, t^{*}$ which satisfy the equality $\sqrt{\frac{s^{*}}{t^{*}}} = \iprod{\phi}{\psi}$. $\ket{\phi}$ and $\ket{\psi}$ is then expanded as Eq.~(\ref{eq:pipe}) and we have
\begin{equation}
\sqrt{\frac{s^{*}}{t^{*}}} s_i \ge \sqrt{s_it_i} \iprod{\eta_i}{\gamma_i}.
\end{equation}
Since
\begin{equation}
\sum_{i=m+1}^N \sqrt{\frac{s^{*}}{t^{*}}} s_i \ge \sum_{i=m+1}^N \sqrt{s_it_i} \iprod{\eta_i}{\gamma_i} = \iprod{\phi}{\psi}
\end{equation}
We have
\begin{equation}
\sum_{i=m+1}^N s_i = 1,
\end{equation}
and
\begin{equation}
\sqrt{\frac{s^{*}s_i}{t^{*}t_i}} = \iprod{\eta_i}{\gamma_i}.
\end{equation}
Thus, the success probability is given as
\begin{eqnarray}
P_{max}^{LOCC} & = & \sum_{i=1}^m P_i + \sum_{i=m+1}^N P_i P_{\ket{\psi}|i} (1-\iprod{\eta_i}{\gamma_i}^2) \nonumber\\
               & = & \sum_{i=1}^m (ss_i+tt_i) + \sum_{i=m+1}^N tt_i ( 1 - \frac{s^{*}s_i}{t^{*}t_i} ) \nonumber\\
               & = & t-t\iprod{\phi}{\psi}^2 \nonumber\\
               & = & P_{max}.
\end{eqnarray}

Discrimination of two \textit{multiparite} states with arbitrary prior probability can also be optimally achieved since our protocol is a ``one way'' protocol and the generalization argument used in Ref.~\cite{CY02} also works here.

As our protocol achieves the globally optimal success probability, it is obviously optimal in the restricted operations of LOCC. Further more, this result completely affirms the conjecture that LOCC can perform as well as global operations in conclusive discrimination of any two states with arbitrary prior probability and in some sense indicates that there is no ``non-locality'' in conclusive discrimination of two states.

Yet, it is still too rush to say that no ``non-locality'' exists in discrimination two states according to any figure of merit \cite{VSPM01}. There might be some special tasks which can be regarded as ``discrimination procedures'' and can reveal the ``non-locality'' property in the two states. Thus, at the present stage, this kind of vagary ``discrimination procedures'' deserve further investigation in further researches.

\bibliography{ocdal}

\begin{thebibliography}{10}
\expandafter\ifx\csname natexlab\endcsname\relax\def\natexlab#1{#1}\fi
\expandafter\ifx\csname bibnamefont\endcsname\relax
  \def\bibnamefont#1{#1}\fi
\expandafter\ifx\csname bibfnamefont\endcsname\relax
  \def\bibfnamefont#1{#1}\fi
\expandafter\ifx\csname citenamefont\endcsname\relax
  \def\citenamefont#1{#1}\fi
\expandafter\ifx\csname url\endcsname\relax
  \def\url#1{\texttt{#1}}\fi
\expandafter\ifx\csname urlprefix\endcsname\relax\def\urlprefix{URL }\fi
\providecommand{\bibinfo}[2]{#2}
\providecommand{\eprint}[2][]{\url{#2}}

\bibitem[{\citenamefont{Bennett et~al.}(1999)\citenamefont{Bennett, DiVincenzo,
  Fuchs, Mor, Rains, Shor, Smolin, and Wootters}}]{BDF+99}
\bibinfo{author}{\bibfnamefont{C.~H.} \bibnamefont{Bennett}},
  \bibinfo{author}{\bibfnamefont{D.~P.} \bibnamefont{DiVincenzo}},
  \bibinfo{author}{\bibfnamefont{C.~A.} \bibnamefont{Fuchs}},
  \bibinfo{author}{\bibfnamefont{T.}~\bibnamefont{Mor}},
  \bibinfo{author}{\bibfnamefont{E.}~\bibnamefont{Rains}},
  \bibinfo{author}{\bibfnamefont{P.~W.} \bibnamefont{Shor}},
  \bibinfo{author}{\bibfnamefont{J.~A.} \bibnamefont{Smolin}},
  \bibnamefont{and} \bibinfo{author}{\bibfnamefont{W.~K.}
  \bibnamefont{Wootters}}, \bibinfo{journal}{Physical Review A (Atomic,
  Molecular, and Optical Physics)} \textbf{\bibinfo{volume}{59}},
  \bibinfo{pages}{1070} (\bibinfo{year}{1999}),
  \urlprefix\url{http://link.aps.org/abstract/PRA/v59/p1070}.

\bibitem[{\citenamefont{Walgate et~al.}(2000)\citenamefont{Walgate, Short,
  Hardy, and Vedral}}]{WSHV00}
\bibinfo{author}{\bibfnamefont{J.}~\bibnamefont{Walgate}},
  \bibinfo{author}{\bibfnamefont{A.~J.} \bibnamefont{Short}},
  \bibinfo{author}{\bibfnamefont{L.}~\bibnamefont{Hardy}}, \bibnamefont{and}
  \bibinfo{author}{\bibfnamefont{V.}~\bibnamefont{Vedral}},
  \bibinfo{journal}{Physical Review Letters} \textbf{\bibinfo{volume}{85}},
  \bibinfo{pages}{4972} (\bibinfo{year}{2000}),
  \urlprefix\url{http://link.aps.org/abstract/PRL/v85/p4972}.

\bibitem[{\citenamefont{Virmani et~al.}(2001)\citenamefont{Virmani, Sacchi,
  Plenio, and Markham}}]{VSPM01}
\bibinfo{author}{\bibfnamefont{S.}~\bibnamefont{Virmani}},
  \bibinfo{author}{\bibfnamefont{M.~F.} \bibnamefont{Sacchi}},
  \bibinfo{author}{\bibfnamefont{M.~B.} \bibnamefont{Plenio}},
  \bibnamefont{and} \bibinfo{author}{\bibfnamefont{D.}~\bibnamefont{Markham}},
  \bibinfo{journal}{Physics Letters A} \textbf{\bibinfo{volume}{288}},
  \bibinfo{pages}{62} (\bibinfo{year}{2001}).

\bibitem[{\citenamefont{Chen and Yang}(2001)}]{CY01}
\bibinfo{author}{\bibfnamefont{Y.-X.} \bibnamefont{Chen}} \bibnamefont{and}
  \bibinfo{author}{\bibfnamefont{D.}~\bibnamefont{Yang}},
  \bibinfo{journal}{Physical Review A (Atomic, Molecular, and Optical Physics)}
  \textbf{\bibinfo{volume}{64}}, \bibinfo{eid}{064303}
  (pages~\bibinfo{numpages}{3}) (\bibinfo{year}{2001}),
  \urlprefix\url{http://link.aps.org/abstract/PRA/v64/e064303}.

\bibitem[{\citenamefont{Chen and Yang}(2002)}]{CY02}
\bibinfo{author}{\bibfnamefont{Y.-X.} \bibnamefont{Chen}} \bibnamefont{and}
  \bibinfo{author}{\bibfnamefont{D.}~\bibnamefont{Yang}},
  \bibinfo{journal}{Physical Review A (Atomic, Molecular, and Optical Physics)}
  \textbf{\bibinfo{volume}{65}}, \bibinfo{eid}{022320}
  (pages~\bibinfo{numpages}{4}) (\bibinfo{year}{2002}),
  \urlprefix\url{http://link.aps.org/abstract/PRA/v65/e022320}.

\bibitem[{\citenamefont{Ivanovic}(1987)}]{Ivan87}
\bibinfo{author}{\bibfnamefont{I.~D.} \bibnamefont{Ivanovic}},
  \bibinfo{journal}{Physics Letters A} \textbf{\bibinfo{volume}{123}},
  \bibinfo{pages}{257} (\bibinfo{year}{1987}).

\bibitem[{\citenamefont{Dieks}(1988)}]{Diek88}
\bibinfo{author}{\bibfnamefont{D.}~\bibnamefont{Dieks}},
  \bibinfo{journal}{Physics Letters A} \textbf{\bibinfo{volume}{126}},
  \bibinfo{pages}{303} (\bibinfo{year}{1988}).

\bibitem[{\citenamefont{Peres}(1988)}]{Pere88}
\bibinfo{author}{\bibfnamefont{A.}~\bibnamefont{Peres}},
  \bibinfo{journal}{Physics Letters A} \textbf{\bibinfo{volume}{128}},
  \bibinfo{pages}{19} (\bibinfo{year}{1988}).

\bibitem[{\citenamefont{Jaeger and Shimony}(1995)}]{JS95}
\bibinfo{author}{\bibfnamefont{G.}~\bibnamefont{Jaeger}} \bibnamefont{and}
  \bibinfo{author}{\bibfnamefont{A.}~\bibnamefont{Shimony}},
  \bibinfo{journal}{Physics Letters A} \textbf{\bibinfo{volume}{197}},
  \bibinfo{pages}{83} (\bibinfo{year}{1995}).

\bibitem[{\citenamefont{Hughston et~al.}(1993)\citenamefont{Hughston, Jozsa,
  and Wootters}}]{HJW93}
\bibinfo{author}{\bibfnamefont{L.~P.} \bibnamefont{Hughston}},
  \bibinfo{author}{\bibfnamefont{R.}~\bibnamefont{Jozsa}}, \bibnamefont{and}
  \bibinfo{author}{\bibfnamefont{W.~K.} \bibnamefont{Wootters}},
  \bibinfo{journal}{Physics Letters A} \textbf{\bibinfo{volume}{183}},
  \bibinfo{pages}{14} (\bibinfo{year}{1993}).

\end{thebibliography}

\end{document}